\documentclass[epj]{svjour}
\usepackage[english]{babel}
\usepackage[bookmarks,pdfhighlight=/O,colorlinks=false,pdfstartview=FitH]{hyperref}
\usepackage{amsmath,amssymb,mathrsfs}
\usepackage[dvipsnames]{xcolor}
\usepackage{graphicx}
\usepackage{epstopdf}
\usepackage{multirow}

\DeclareMathOperator\artanh{Artanh}

\begin{document}
	\title{Probing chemical freeze-out criteria in relativistic nuclear collisions with coarse grained transport simulations}
	\author{Tom Reichert\inst{1,2} \and Gabriele Inghirami\inst{3,4} \and Marcus Bleicher\inst{1,2,5,6}
	}                     
	%
	%
	\institute{Institut f\"ur Theoretische Physik, Goethe Universit\"at Frankfurt, Max-von-Laue-Str. 1, D-60438 Frankfurt am Main, Germany \and Helmholtz Research Academy Hesse for FAIR, Campus Frankfurt, Max-von-Laue-Str. 12, 60438 Frankfurt,  Germany \and University of Jyv\"askyl\"a,  Department of Physics, P.O. Box 35, FI-40014 University of Jyv\"askyl\"a, Finland \and Helsinki Institute of Physics, P.O. Box 64, FI-00014 University of Helsinki, Finland \and GSI Helmholtzzentrum f\"ur Schwerionenforschung GmbH, Planckstr. 1, 64291 Darmstadt , Germany \and John von Neumann-Institut f\"ur Computing, Forschungszentrum J\"ulich, 52425 J\"ulich, Germany}
	\date{Received: \today / Revised version: }
	%
	\abstract{
		We introduce a novel approach based on elastic and inelastic scattering rates to extract the hyper-surface of the chemical freeze-out from a hadronic transport model in the energy range from E$_\mathrm{lab}=1.23$~AGeV to $\sqrt{s_\mathrm{NN}}=62.4$~GeV. For this study, the Ultra-relativistic Quantum Molecular Dynamics (UrQMD) model combined with a coarse-graining method is employed. The chemical freeze-out distribution is reconstructed from the pions through several decay and re-formation chains involving resonances and taking into account inelastic, pseudo-elastic and string excitation reactions. The extracted average temperature and baryon chemical potential are then compared to statistical model analysis. Finally we investigate various freeze-out criteria suggested in the literature. We confirm within this microscopic dynamical simulation, that the chemical freeze-out at all energies coincides with $\langle E\rangle/\langle N\rangle\approx1$~GeV, while other criteria, like $s/T^3=7$ and $n_\mathrm{B}+n_\mathrm{\bar{B}}\approx0.12$ fm$^{-3}$ are limited to higher collision energies.
	} 
	\authorrunning{T.~Reichert~\textit{et al.}}
	\titlerunning{Probing chemical f.o. criteria in rel. nuclear collisions with coarse grained transport simulations}
	\maketitle
\section{Introduction}\label{intro}
The collision of heavy ions in today's largest particle accelerators provides an excellent tool to explore nuclear and sub-nuclear matter under extreme conditions as they occur e.g. in neutron stars, around black holes or in the early universe. Matter created under these conditions sustains tremendous temperatures, pressures and densities in volumes on the order of $\sim$~1000~fm$^3$ over timescales of $10^{-23}$~s.

Since separating quarks creates new quark-anti-quark pairs from the vacuum in order to bind to color neutral hadrons, a direct measurement of the inner degrees of freedom of strongly interacting matter in heavy ion collisions is, unfortunately, still impossible. In contrast to QED, this phenomenon called confinement forbids perturbative calculations at small momenta. Access to the early and intermediate stage of a collision can be gained via electromagnetic probes e.g. with real photons and virtual photons in the di-lepton channel \cite{Feinberg:1976ua,Shuryak:1978ij,Bratkovskaya:2014mva}, via the study of hadrons produced at the chemical freeze-out \cite{Rafelski:1982pu,BraunMunzinger:2003zz}, or by flow observables, like $v_1$, $v_2$, $v_3$, $\cdots$. 

Typically, the final state of a heavy ion collision involves two parts: 1.) the chemical freeze-out and 2.) the kinetic (or thermal) freeze-out. At the chemical freeze-out inelastic flavour changing reactions cease (particle yields get fixed) and at the kinetic freeze-out also elastic reactions cease (particle 4-momenta get fixed) and the system decouples. This behavior is also reflected in the scattering rates as shown in \cite{Bleicher:2002dm} where the chemical freeze-out is estimated to occur at $\tau_\mathrm{chem}=6$~fm. It has been shown, that the kinetic freeze-out proceeds in a more continuous fashion \cite{Inghirami:2019muf} than being an instantaneous process. Assuming thermal equilibrium, particle spectra or ratios can be fitted with statistical and blastwave models to extract kinetic or chemical freeze-out properties like the temperature $T$ and the baryo-chemical potential $\mu_\mathrm{B}$. Thermal fits of particle ratios resulting from CERN/SPS \cite{BraunMunzinger:1995bp,BraunMunzinger:1999qy,Letessier:1998ca,Spieles:1997tf}, BNL/AGS \cite{Letessier:1994cn,Becattini:1998zd,Cleymans:1996cd} and GSI/SIS \cite{Cleymans:1997sw,Cleymans:1998yf} measurements led to a unified description of the chemical freeze-out line via a hadronic gas model at a constant energy per particle of $\langle E\rangle/\langle N\rangle=1$~GeV \cite{Cleymans:1998fq}. Besides this, other quantities were proposed to define the chemical freeze-out, e.g. a constant total baryon density $n_\mathrm{B}+n_\mathrm{\bar{B}}=0.12$~fm$^{-3}$ \cite{Braun_Munzinger_2002} or a constant entropy per temperature $s/T^3=7$ \cite{CLEYMANS200550}.

The present investigation introduces a novel approach to determine the chemical freeze-out hyper-surface directly from a combined UrQMD/coarse-grained framework. The full time evolution of Au+Au collisions in the GSI/FAIR and RHIC BES-II energy regime is analyzed for this purpose and every final state pion is traced back to the space-time point of its creation taking into account absorption and decay processes as e.g. $N+\pi\leftrightarrow\Delta$ which affect the freeze-out coordinates. These space-time points define the hyper-surface of the chemical freeze-out of the pions. We focus on pions, because they are the most abundant hadrons and they are produced in sufficient amount at all investigated energies. To obtain the thermodynamic parameters ($T$, $\mu_\mathrm{B}$) on this hyper-surface, the UrQMD data is coarse grained and supplemented by a Hadron Resonance Gas EoS \cite{Zschiesche:2002zr}. The energy dependence of the calculated thermodynamic variables is then investigated to test different suggested criteria for the chemical freeze-out.

\section{The model}\label{sec:procedures}
The present study uses the Ultra-relativistic Quantum Molecular Dynamics (UrQMD)~\cite{Bass:1998ca,Bleicher:1999xi} transport model in cascade mode. UrQMD is used to compute both the bulk evolution of the system and to determine the space-time coordinates of the chemical freeze-out of the hadrons. We recall that UrQMD is a hadron cascade model that simulates the dynamical evolution of heavy ion collision events by following the propagation of the individual hadrons, modeling their interactions via the excitation of color flux-tubes (strings) and by further elastic and inelastic scatterings. A transition to a deconfined stage is not explicitly included in the cascade mode employed here.

\subsection{The coarse graining approach}
The UrQMD coarse-graining approach~\cite{Inghirami:2019muf,Huovinen:2002im,Endres:2014zua,Endres:2015fna,Endres:2015egk,Inghirami:2018vqd} consists in computing the temperature and the baryon chemical potential from the average energy-momentum tensor and net baryon current of the hadrons formed in a large set of heavy ion collision events with the same collision energy and centrality. The computation is done in the cells of a fixed spatial grid at constant intervals of time. In the present study, the cells are four-cubes with spatial sides of length $\Delta x=\Delta y=\Delta z=1$~fm and $\Delta t=0.25$~fm length in time direction. First, we evaluate the net-baryon four current $j^{\mu}_{\mathrm{B}}$ as
\begin{equation}
j^{\mu}_{\mathrm{B}}(t,\vec{r})=\frac{1}{\mathrm{\Delta} V}\left\langle \sum\limits_{i=1}^{N_h \in \Delta V} B_i \frac{p^{\mu}_{i}}{p^{0}_{i}}\right\rangle,
\label{eq:j_avg}
\end{equation}
and the energy momentum tensor $T^{\mu\nu}$ as
\begin{equation}
T^{\mu\nu}(t,\vec{r})=\frac{1}{\Delta V}\left\langle \sum\limits_{i=1}^{N_{h} \in \Delta V} \frac{p^{\mu}_{i} p^{\nu}_{i}}{p^{0}_{i}}\right\rangle,
\label{eq:Tmunu_avg}
\end{equation}
in which $\Delta V$ is the volume of the cell, $B_i$ is the baryon number and $p^{\mu}_i$ stands for the $\mu$ component of the four momentum of the hadron $i$. The sums run over all hadrons $N_{h}$ in the cell. We adopt the Eckart's frame definition~\cite{Eckart:1940te} and we obtain the fluid four velocity $u^{\mu}$ from $j^{\mu}_{\mathrm{B}}$ as
\begin{equation}
u^{\mu}=\dfrac{j^{\mu}_{\mathrm{B}}}{\sqrt{j^{\nu}_{\mathrm{B}}j^{}_{\mathrm{B}\nu}}}=(\gamma,\gamma \vec{v}),
\label{eq:fluid_vel}
\end{equation}
in which $\gamma$ is the Lorentz factor and $v$ the fluid velocity in natural units ($c=\hbar=1$). By a Lorentz transformation of the net-baryon current and of the energy momentum tensor in the Local Rest Frame (LRF) of the fluid, we compute the baryon density $\rho_{\mathrm{B}}$ and the energy density $ \varepsilon$ as:
\begin{equation} 
\rho_{\mathrm{B}} = j_{\mathrm{B,\,LRF}}^{0},\qquad \varepsilon = T^{00}_{\mathrm{LRF}}.
\end{equation} 
Often the chemical freeze-out occurs in cells with an an\-isotropy between the pressure in the parallel ($P_{\parallel}$) and in the transverse direction ($P_{\perp}$), with respect to the beam axis. To take into account this condition, we rescale $\varepsilon$~\cite{Florkowski:2010cf,Ryblewski:2012rr,Endres:2014zua,Moreau:2019vhw} as:
\begin{equation}
\varepsilon_{corr}=\varepsilon/r(\chi),
\label{eq:energy_dens_rescaling}
\end{equation}
where $\chi=(P_{\perp}/P_{\parallel})^{4/3}$ and 
\begin{equation}
r(\chi) =
\left\{
\begin{aligned}
\frac{\chi^{-1/3}}{2} \left[1+\frac{\chi \artanh \sqrt{1-\chi}}{\sqrt{1-\chi}}\right],\, \mathrm{if}\; \chi < 1\\
\frac{\chi^{-1/3}}{2} \left[1+\frac{\chi \artanh \sqrt{\chi-1}}{\sqrt{\chi-1}}\right], \, \mathrm{if}\; \chi > 1
\end{aligned}
\right.
.
\label{eq:r_correction}
\end{equation}
This corrections was also applied in all our previous studies~\cite{Endres:2014zua,Endres:2015fna,Endres:2015egk,Inghirami:2018vqd}.
The final step in the coarse graining procedure consists in associating to each cell of the coarse grained grid the temperature $T(\varepsilon_{corr},\rho_{\mathrm{B}})$ and the baryon chemical potential $\mu_{\mathrm{B}}(\varepsilon_{corr},\rho_{\mathrm{B}})$ through the interpolation of a tabulated Hadron Resonance Gas EoS~\cite{Zschiesche:2002zr}.

\subsection{Chemical freeze-out in the UrQMD model}\label{sec:2.1}
To test this novel way of tracking down the chemical freeze-out coordinates in a full transport simulation we focus on pions. The reason for this is twofold: a) pions are very abundant hadrons in the investigated energy regime and b) pions are stable particles under strong interactions, both facts simplify the reconstruction of the chemical freeze-out coordinates with high accuracy. When does a $\pi$ freeze-out chemically? Typically, pions are either produced directly in a string decay (dominant at higher energies) or via $N+N\rightarrow N+\Delta$, and subsequently $\Delta\rightarrow N+\pi$ reactions. Of course the $\Delta$ can be replaced by other resonances. In addition cascades like $\Delta\rightarrow N+\rho$, and subsequently $\rho\rightarrow\pi\pi$ are possible. Not all pions produced initially make it to the final state due to absorption processes, e.g. $\pi+N\rightarrow N^*\rightarrow K+\Lambda$. To extract the space-time point of the production of a finally observed $\pi$, we follow all observed pions backwards through the evolution until we reach their point of production (i.e. the point at which either their mother-resonances were formed or the pions were produced directly e.g. from a string). This defines the chemical freeze-out coordinates $(t,\vec{r})$ for each individual pion. 

\section{Results}
The results are obtained by analyzing central Au+Au collisions calculated with the UrQMD model. The simulations are used as input to extract the chemical freeze-out coordinates and calculate the fields $(T(t,\vec{r}),\mu_\mathrm{B}(t,\vec{r}))$ with the coarse-graining procedure. Central events are selected via an impact parameter cut at $b_\mathrm{max}=3.4$~fm at all investigated energies. We focus on the chemical freeze-out of pions which finally decouple in the volume with $|z|\leq 5$~fm. 

\subsection{Chemical freeze-out times}\label{sec:3-2}
\begin{figure} [t!hb]
	\resizebox{0.5\textwidth}{!}{
		\includegraphics{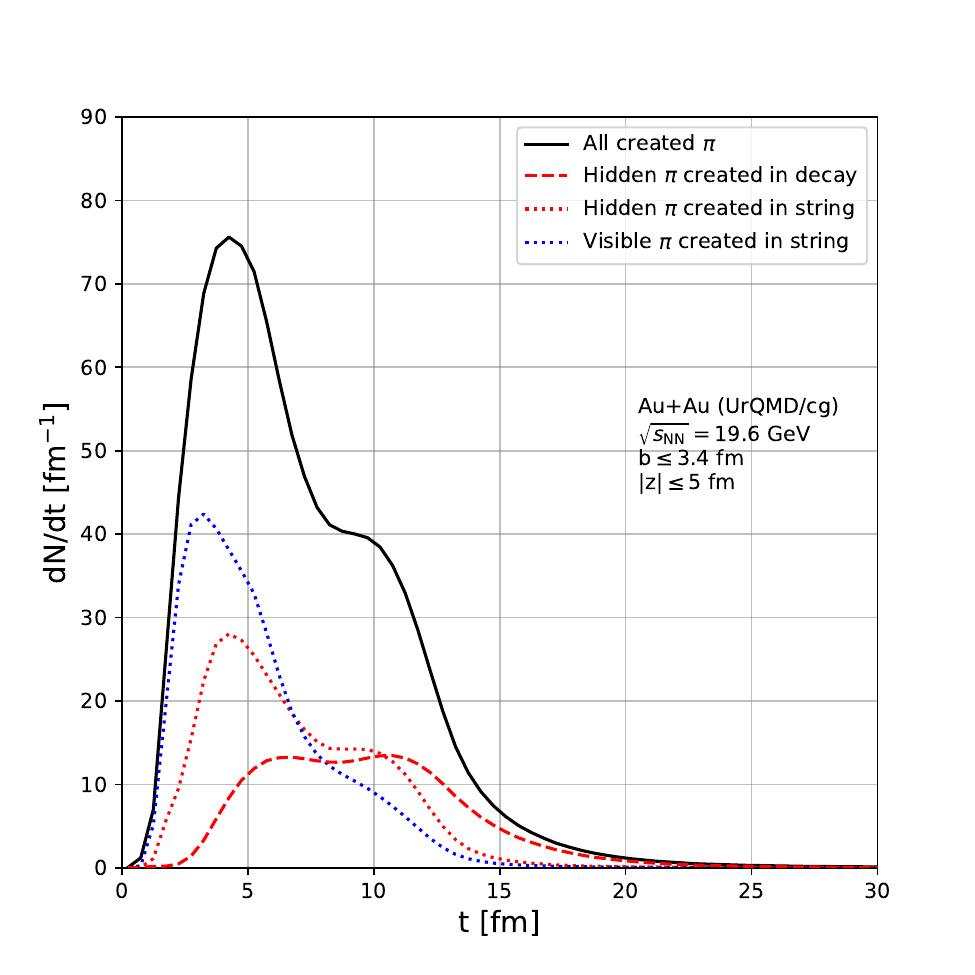}
	}
	\caption{[Color online] Contributions to the full chemical freeze-out distribution of $\pi$ having their last interaction in $|z|\leq5$~fm and created in central Au+Au collisions in the UrQMD model. Hidden $\pi$ are shown in red and visible $\pi$ are shown in blue, while dashed lines represent decays and dotted lines stand for string reactions.}
	\label{dndt_chem_contributions}
\end{figure}

\begin{figure} [t!hb]
	\resizebox{0.5\textwidth}{!}{
		\includegraphics{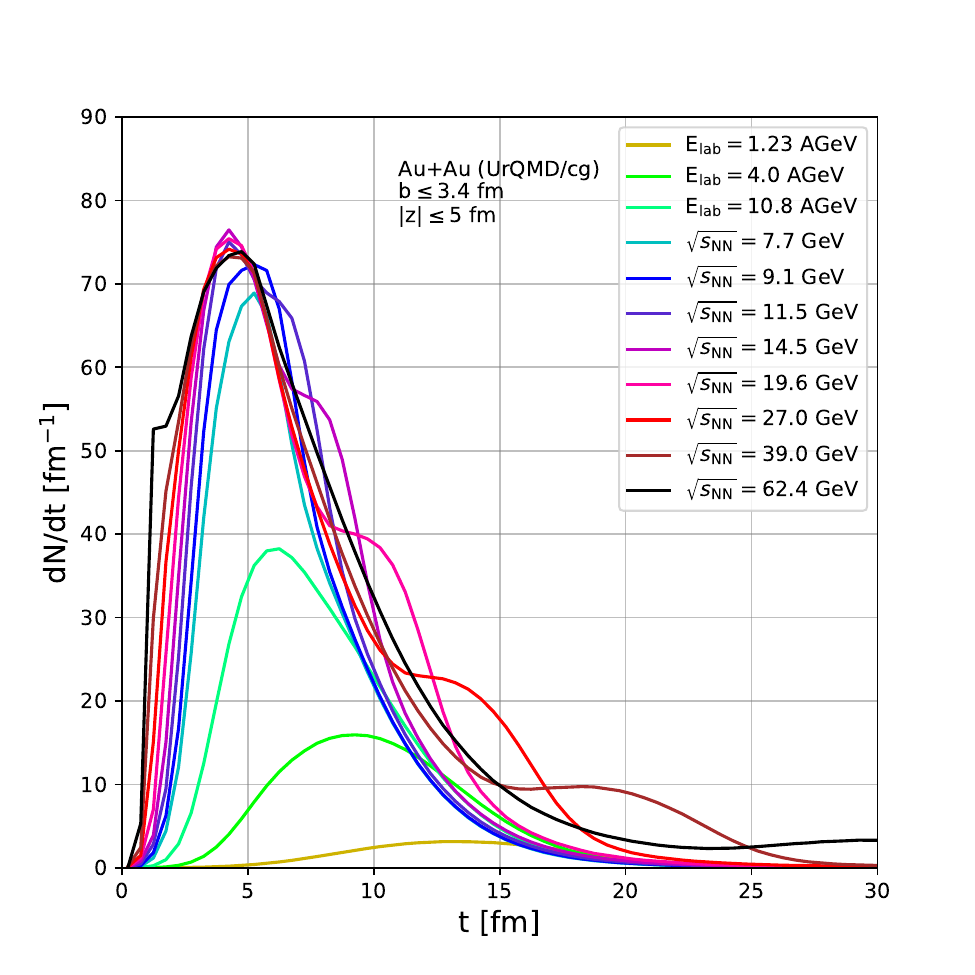}
	}
	\caption{[Color online] Distribution of the chemical freeze-out times of $\pi$'s reconstructed with the presented algorithm at $|z|\leq5$~fm at E$_\mathrm{lab}=1.23$~AGeV (yellow), E$_\mathrm{lab}=4.0$~AGeV (lime), E$_\mathrm{lab}=10.8$~AGeV (springgreen), $\sqrt{s_\mathrm{NN}}=$ 7.7~GeV (cyan), $\sqrt{s_\mathrm{NN}}=$ 9.1~GeV (blue), $\sqrt{s_\mathrm{NN}}=$ 11.5~GeV (magenta), $\sqrt{s_\mathrm{NN}}=$ 14.5~GeV (purple), $\sqrt{s_\mathrm{NN}}=$ 19.6~GeV (pink), $\sqrt{s_\mathrm{NN}}=$ 27~GeV (red), $\sqrt{s_\mathrm{NN}}=$ 39~GeV (brown) and $\sqrt{s_\mathrm{NN}}=$ 62.4~GeV (black) in central Au+Au collisions ($b\leq3.4$~fm) from UrQMD.}
	\label{dndt_chem}
\end{figure}

Let us start with the distribution of the chemical freeze-out times at a specific energy to first analyze the influence of the different reconstruction scenarios. For this purpose we show the contributions to the full chemical freeze-out distribution at $\sqrt{s_\mathrm{NN}}=19.6$~GeV in Fig. \ref{dndt_chem_contributions}. The total reconstructed $\pi$ distribution is shown as a solid black line. It consists of pions that are created hidden in carriers (e.g. $\Delta$, $\rho$, etc.), shown as red lines and those created directly as pions, shown as blue line. In general the full distribution shows two features: one local maximum centered at $\approx$~4~fm and a second small bump arising at $\approx$~10~fm. The first peak consists mainly of hadrons created in string excitation reactions, while the second bump consists of decays of resonances and secondary strings. 
To better interpret the appearing structure, we now show in Fig. \ref{dndt_chem} the chemical freeze-out distributions of $\pi$ mesons reconstructed with the presented algorithm at $|z|\leq5$~fm at the collision energies E$_\mathrm{lab}=1.23$~AGeV (yellow), E$_\mathrm{lab}=4.0$~AGeV (lime), E$_\mathrm{lab}=10.8$~AGeV (springgreen), $\sqrt{s_\mathrm{NN}}=$ 7.7~GeV (cyan), $\sqrt{s_\mathrm{NN}}=$ 9.1~GeV (blue), $\sqrt{s_\mathrm{NN}}=$ 11.5~GeV (magenta), $\sqrt{s_\mathrm{NN}}=$ 14.5~GeV (purple), $\sqrt{s_\mathrm{NN}}=$ 19.6~GeV (pink), $\sqrt{s_\mathrm{NN}}=$ 27~GeV (red), $\sqrt{s_\mathrm{NN}}=$ 39~GeV (brown) and $\sqrt{s_\mathrm{NN}}=$ 62.4~GeV (black) in central Au+Au collisions ($b\leq3.4$~fm) from UrQMD. All distributions are normalized to the number of events. The curves corresponding to the RHIC BES-II energies generally show the same structure: First, a strongly emphasized peak centered at $\approx$~5-10~fm (depending on the energy) and second a small bump which shifts towards later times with increasing energy, which is however not visible in the GSI/FAIR energy regime. Here, only the pronounced maximum appears. It is interesting that the peak of the body of the distribution does only very weakly vary with the collision energy in the explored energy regime. Pions created here will mainly evolve through several decay and re-formation cycles of short lived resonances like the $\Delta$ or the $\rho$. When comparing our chemical freeze-out time distribution with the one given in \cite{Inghirami:2019muf} where the authors investigated the kinetic freeze-out, we clearly observe that the chemical freeze-out exhibits a more narrow distribution than the kinetic freeze-out. This finding validates therefore that the conceptual idea of an instantaneous chemical freeze-out is to some extent reasonable while the kinetic freeze-out studied in e.g. \cite{Inghirami:2019muf} is shown to be a more continuous and spread-out process. The position of the small bump in contrast depends on the energy. We relate this effect to the $\gamma$ factor of resonances formed in string decays. If e.g. a $\Delta$ resonance is formed from a di-quark within a string, which decays into a $\rho+N$ and subsequently into a final state pion, the creation time satisfies: $t_\mathrm{chem.}\approx\sqrt{s_\mathrm{NN}}/(2m_p)+\Gamma_\Delta^{-1}$.
The average chemical freeze-out times $\tau_\mathrm{chem}$ can be found in Tab. \ref{summary_table} stating that the chemical freeze-out in the RHIC BES-II energy regime occurs at $\approx$~7~fm.

\subsection{Temperature and chemical potential distributions}\label{sec:3-3}
\begin{figure} [t!hb]
	\resizebox{0.5\textwidth}{!}{
		\includegraphics{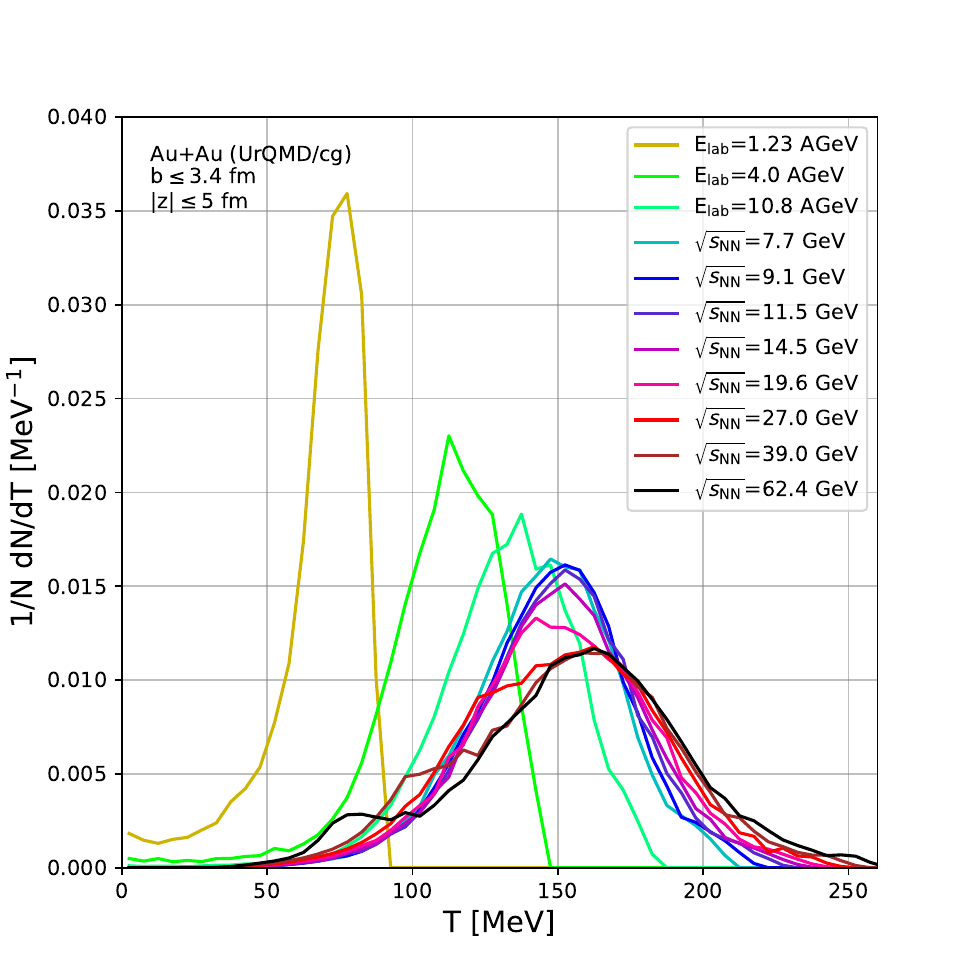}
	}
	\caption{[Color online] Distribution of the chemical freeze-out temperatures of $\pi$'s reconstructed with the presented algorithm at $|z|\leq5$~fm from E$_\mathrm{lab}=1.23$~AGeV to $\sqrt{s_\mathrm{NN}}=$ 62.4~GeV (same colors as in Fig. \ref{dndt_chem}) in central Au+Au collisions ($b\leq3.4$~fm) from UrQMD.}
	\label{dndT_chem}
\end{figure}

\begin{figure} [t!hb]
	\resizebox{0.5\textwidth}{!}{
		\includegraphics{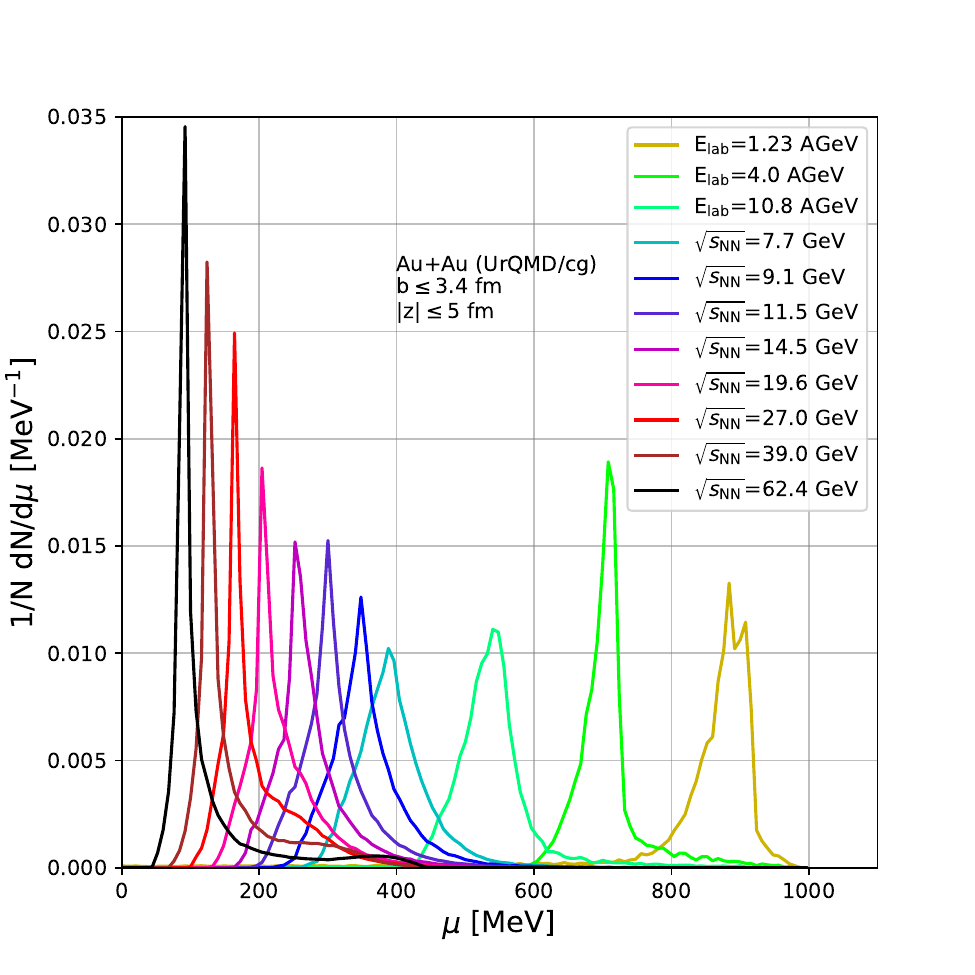}
	}
	\caption{[Color online] Distribution of the baryon chemical potential at the chemical freeze-out of $\pi$'s reconstructed with the presented algorithm at $|z|\leq5$~fm from E$_\mathrm{lab}=1.23$~AGeV to $\sqrt{s_\mathrm{NN}}=$ 62.4~GeV (same colors as in Fig. \ref{dndt_chem}) in central Au+Au collisions ($b\leq3.4$~fm) from UrQMD.}
	\label{dndmu_chem}
\end{figure}

Now that the freeze-out four-coordinates are determined, the question arises how the coarse-grained temperatures and the baryon chemical potentials will be distributed. For this we show in Fig. \ref{dndT_chem} the distribution of the coarse-grained chemical freeze-out temperatures of pions with $|z|\leq5$~fm in Au+Au collisions from E$_\mathrm{lab}=1.23$~AGeV to $\sqrt{s_\mathrm{NN}}=$ 62.4~GeV (same colors as in Fig. \ref{dndt_chem}) from UrQMD as well as we show the distributions of the baryon chemical potential at the chemical freeze-out in Fig. \ref{dndmu_chem}.
We observe two major features in the distributions of temperatures extracted from the reconstructed hypersurface: a) with increasing beam energy, the temperature distributions merge into one peak at $\approx$~150~MeV (in line with expectations from a thermal model analysis) while the typical widths of the temperature spread is FWHM $\approx50$~MeV, different from the expectations from a thermal model analysis with a single temperature. 
b) The distributions of the baryon chemical potentials at chemical freeze-out is very narrow at each energy. As expected the chemical potential decreases with increasing energy. In contrast to the temperatures we do not observe a saturation of $\mu_\mathrm{B}$, but a continuous decrease.

\subsection{Energy dependence}
\begin{figure} [t!hb]
	\resizebox{0.5\textwidth}{!}{
		\includegraphics{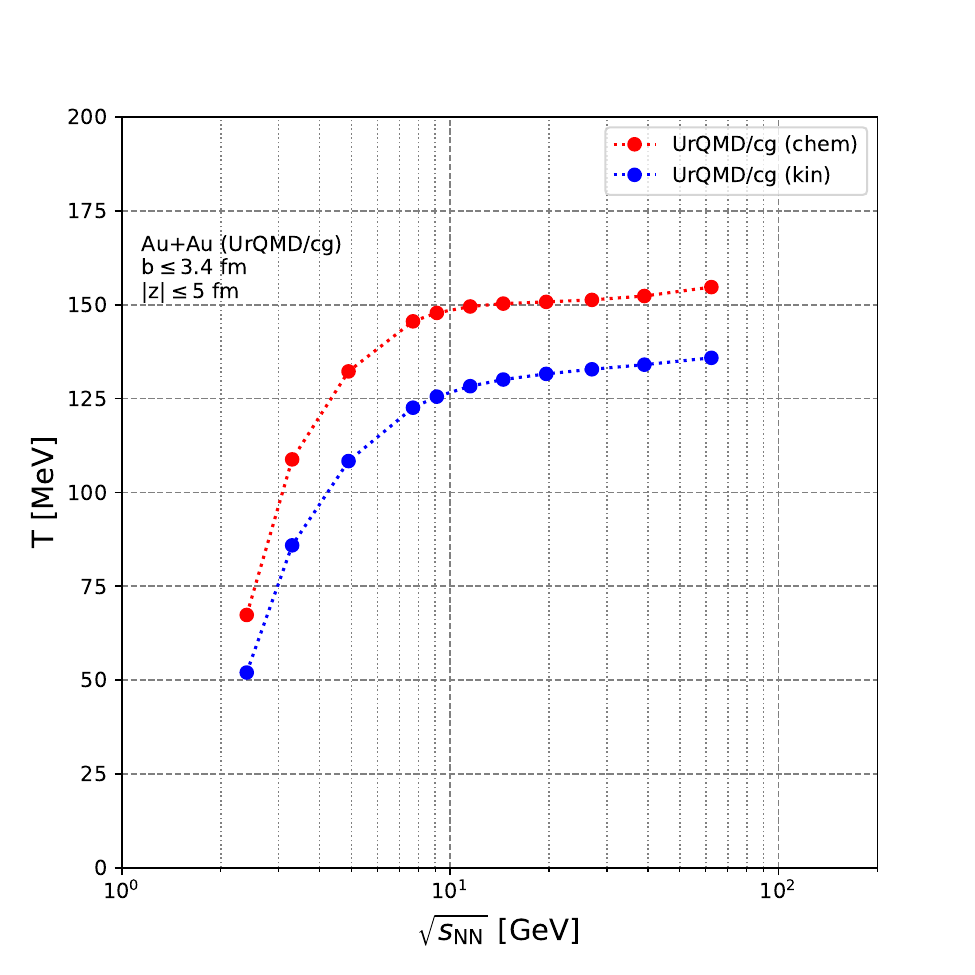}
	}
	\caption{[Color online] Average temperature at the chemical freeze-out (red circles) and at the kinetic freeze-out (blue circles) extracted from the coarse-graining procedure with a HRG EoS in central Au+Au collisions ($b\leq3.4$~fm) from UrQMD.}
	\label{T_sNN}
\end{figure}

\begin{figure} [t!hb]
	\resizebox{0.5\textwidth}{!}{
		\includegraphics{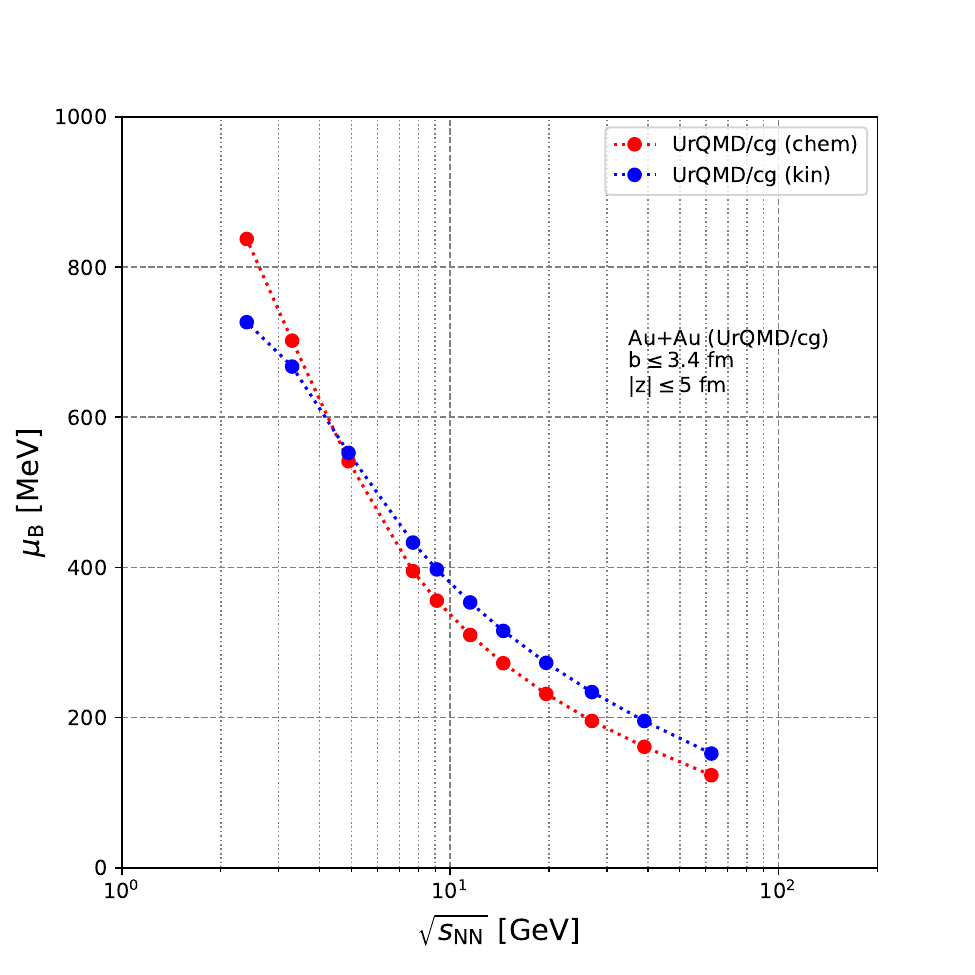}
	}
	\caption{[Color online] Average baryon chemical potential at the chemical freeze-out (red circles) and at the kinetic freeze-out (blue circles) extracted from the coarse-graining procedure with a HRG EoS in central Au+Au collisions ($b\leq3.4$~fm) from UrQMD.}
	\label{mu_sNN}
\end{figure}

In this section the collision energy dependence of the temperature and the baryon chemical potential are investigated on the chemical freeze-out hyper surface. We compare the chemical freeze-out values to kinetic freeze-out values. The kinetic freeze-out is defined as the point of the particles last interaction and can also easily be extracted from the presented algorithm. We show in Fig. \ref{T_sNN} the average temperature at the kinetic freeze-out (blue circles) and the average temperature at the chemical freeze-out (red circles) as a function of the collision energy. Firstly, it is noticeable and important that $T_\mathrm{chem}>T_\mathrm{kin}$ over the whole energy range. This implies that indeed the chemical and the kinetic freeze-out processes happen sequentially. The energy dependence of the absolute difference $\Delta T=T_\mathrm{chem}-T_\mathrm{kin}$ between the two freeze-out temperatures stays according to the present analysis approximately constant ($20\pm5$~MeV). A saturation of the chemical freeze-out temperature is observed at $\approx150$~MeV while the kinetic freeze-out temperature saturates at $\approx130$~MeV.
Fig. \ref{mu_sNN} shows the average baryon chemical potentials at the chemical and the kinetic freeze-out. Both quantities drop rapidly with increasing energy, lining up with the vanishing chemical potential inferred from observed baryon-charge symmetric matter formed at top RHIC and LHC energies. 

\subsection{Phase diagram}\label{sec:3-4}
\begin{figure} [t!hb]
	\resizebox{0.5\textwidth}{!}{
		\includegraphics{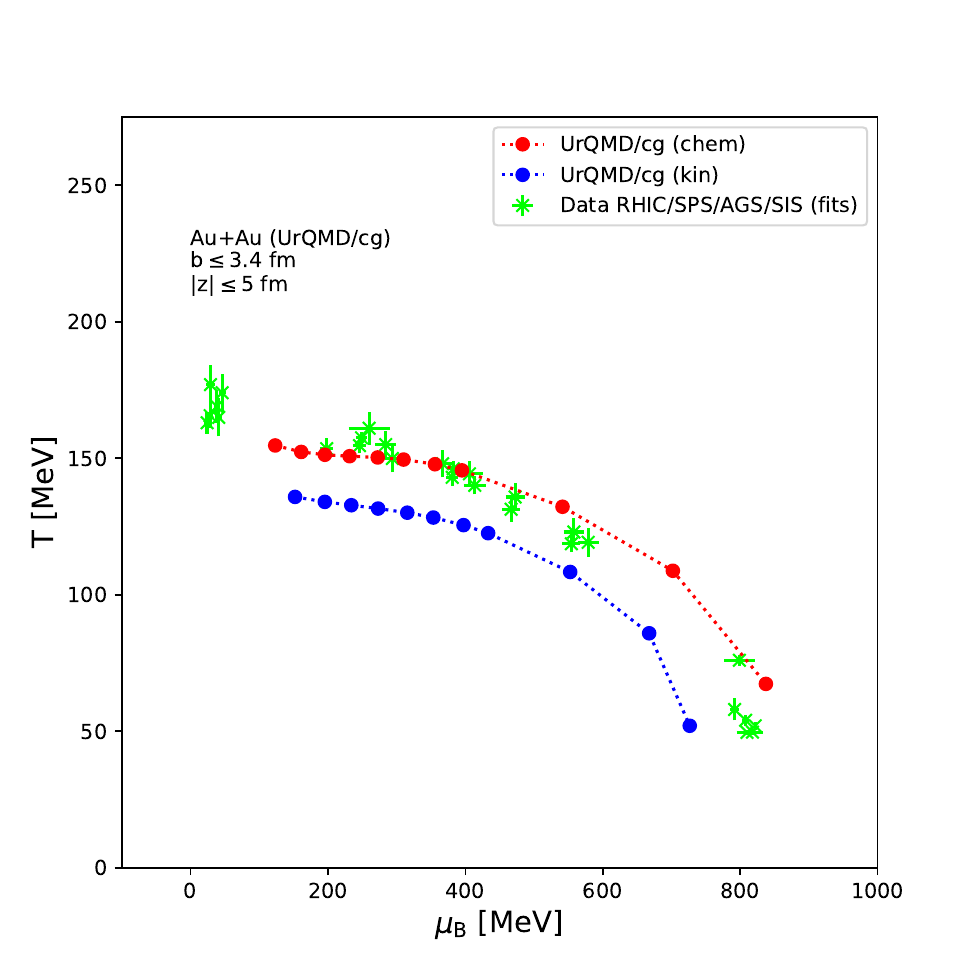}
	}
	\caption{[Color online] Phase diagram of the chemical freeze-out calculated with the UrQMD/cg model (red circles) and of the kinetic freeze-out (blue circles). Also shown are thermal model fits to RHIC \cite{Cleymans:2004pp,BraunMunzinger:2001ip,Adams:2005dq,Florkowski:2001fp,Baran:2003nm}, SPS \cite{Becattini:2003wp,Bravina:2002wz}, AGS \cite{Becattini:2003wp,Bravina:2002wz}, SIS data \cite{Cleymans:1998yb,Becattini:2000jw,Averbeck:2000sn} and the recent HADES point \cite{Agakishiev:2010rs} as green stars.}
	\label{phase_diagram}
\end{figure}

After the analysis of the time distributions and the energy dependence of the chemical freeze-out quantities (temperatures and baryon chemical potentials), we will now relate our results to the phase structure of QCD in the phase diagram of nuclear matter. Fig. \ref{phase_diagram} shows the calculated average temperature $\langle T\rangle$ and the average chemical potential $\langle\mu_{\rm B}\rangle$ as red circles for the chemical freeze-out and as blue circles for the kinetic freeze-out. Fits to the experimental data points obtained at RHIC \cite{Cleymans:2004pp,BraunMunzinger:2001ip,Adams:2005dq,Florkowski:2001fp,Baran:2003nm}, CERN/SPS \cite{Becattini:2003wp,Bravina:2002wz}, BNL/AGS \cite{Becattini:2003wp,Bravina:2002wz}, GSI/SIS \cite{Cleymans:1998yb,Becattini:2000jw,Averbeck:2000sn} and the recent HADES point \cite{Agakishiev:2010rs} are shown as green stars (summarized in \cite{Cleymans:2005xv,Andronic:2005yp}).

We observe in general that our calculated chemical freeze-out line matches the estimates from thermal model fits to the experimental data surprisingly well. This result is remarkable because it is ex-ante not expected that a purely hadronic transport model, which does neither involve the concept of the chemical freeze-out explicitly nor a phase-transition to a deconfined phase, should reproduce ($T$, $\mu_{\rm B}$) combinations near the data points obtained from thermal model fits. This raises a question: Why does a non-equilibrium transport model without phase-transition and chemical break-up reproduce thermal (i.e. equilibrium) quantities such precise? A hint towards the answer can be found the scattering rates. It is well known from e.g. chemistry that the onset of equilibrium can be characterized by the ratio of the expansion rate to the scattering rate. This means to maintain equilibrium, the (inelastic) scattering rate $\Gamma$ must be much larger than the expansion rate $\theta$. Where $\theta/\Gamma=Kn$, can be interpreted as the Knudsen number $Kn$. Typically, $\Gamma\sim f_if_j\sigma_{ij}$, with $f_i$ being the phase space density of species $i$ and $\sigma_{ij}$ being the (inelastic) interaction cross section, and $\theta=\partial_\mu u^\mu$, the divergence of the 4-velocity field \cite{Ahmad:2016ods}. I.e., if the scattering rate is larger than the expansion rate then enough equilibrating reactions happen to keep the system in equilibrium. If otherwise the scattering rate is smaller than the expansion rate then the system does not have enough time to adjust its properties and the state becomes frozen for the remaining time of the expansion. While in our analysis, the competition between scattering rate and expansion rate is modelled by microscopic dynamics, it can also be used to calculate the chemical and kinetic freeze-out in a self consistent way in expanding gases and fluids, see e.g. \cite{Ahmad:2016ods,Blaschke:2017lvd}. 

Therefore, the temperatures and the baryon chemical potentials at chemical and kinetic freeze-out as calculated here emerge from the local interplay of the elastic and inelastic collision rates with the local expansion rates of the system. Thus, the chemical and kinetic freeze-out lines are not related to a phase-transition or a cross over. 

\subsection{Testing chemical freeze-out criteria}
Our analysis can be further deepened by the investigation of quantities which have been proposed to classify the chemical freeze-out line. The systematic investigation of the chemical freeze-out in experiments and theory led to the proposal of several quantities which apparently seem to function as criteria for the chemical freeze-out to occur:
\begin{enumerate}
	\item In Ref. \cite{Cleymans:1998fq} the average energy per particle $\langle E\rangle/\langle N\rangle$ was proposed to stay constant along the chemical freeze-out line with $\langle E\rangle/\langle N\rangle=1$~GeV.
	\item Secondly, an entropy based criterion, $s/T^3=7$ \cite{CLEYMANS200550} was suggested for the meson dominated energy regime.
	\item It was further proposed that the total baryon number can also serve as an estimation for the freeze-out line by demanding $n_\mathrm{B}+n_\mathrm{\bar{B}}=0.12$~fm$^{-3}$ \cite{Braun_Munzinger_2002}.
\end{enumerate}
These suggestions can be directly accessed and tested in the coarse-graining method employed here.

\begin{table*} [t]
	\caption{For each energy (column 1) the average chemical freeze-out time $\tau_\mathrm{chem}$ of all pions (column 2), the average chemical freeze-out temperature $\mathrm{T}_\mathrm{chem}$ (column 3), the average baryon chemical potential at the chemical freeze-out $\mu_\mathrm{B}^\mathrm{chem}$ (column 4) and the freeze-out criteria evaluated at the chemical freeze-out $\langle E\rangle/\langle N\rangle$ (column 5), $s/T^3$ (column 6) and $n_\mathrm{B}+n_\mathrm{\bar{B}}$ (column 7) are shown.}
	\label{summary_table}   
	\resizebox{\textwidth}{!}{
	\begin{tabular}{ccccccc}
		\hline\noalign{\smallskip}
		$\sqrt{s_\mathrm{NN}}$ [GeV] & $\langle\tau_\mathrm{chem}\rangle$ [fm] & $\langle \mathrm{T}_\mathrm{chem}\rangle$ [MeV] & $\langle\mu_\mathrm{B}^\mathrm{chem}\rangle$ [MeV] & $\langle E\rangle/\langle N\rangle$ [GeV] & $s/T^3$ & $n_\mathrm{B}+n_\mathrm{\bar{B}}$ [1/fm$^3$] \\
		\noalign{\smallskip}\hline\noalign{\smallskip}
		2.4 & 13.7 & 67.3 & 837.5 & 1.088 & 31.478 & 0.334 \\
		3.3 & 10.3 & 108.8 & 702.1 & 1.194 & 13.626 & 0.347 \\
		4.9 & 8.1 & 132.2 & 541.5 & 1.166 & 9.248 & 0.303 \\
		7.7 & 6.9 & 145.6 & 395.1 & 1.092 & 7.350 & 0.233 \\
		9.1 & 6.7 & 147.8 & 355.8 & 1.068 & 6.999 & 0.209 \\
		11.5 & 6.7 & 149.5 & 310.1 & 1.038 & 6.607 & 0.178 \\
		14.5 & 6.8 & 150.2 & 272.4 & 1.015 & 6.321 & 0.153 \\
		19.6 & 7.1 & 150.7 & 231.6 & 0.995 & 6.100 & 0.129 \\
		27.0 & 7.5 & 151.3 & 195.5 & 0.987 & 6.047 & 0.111 \\
		39.0 & 8.0 & 152.3 & 161.1 & 0.992 & 6.151 & 0.096 \\
		62.4 & 8.4 & 154.7 & 123.2 & 1.018 & 6.449 & 0.081 \\
		\noalign{\smallskip}\hline
	\end{tabular}
	}
\end{table*}

\subsubsection{Average energy per particle}
With increasing collision energy a major part of the additional energy is used to produce new (heavier) particles than to increase the momentum of the existing particles \cite{Hagedorn:1965st} leading to the limiting Hagedorn temperature for hadrons. In fact, the study of hadronic abundances at SIS, AGS, SPS, RHIC and LHC energies has shown that the chemical freeze-out line can be characterized by a constant $\langle E\rangle/\langle N\rangle$ of $\approx1$~GeV \cite{Cleymans:1998fq} known as the Cleymans-Redlich criterion. This pioneering observation induced many other works to investigate measured hadron ratios which concluded to similar values ranging between $\langle E\rangle/\langle N\rangle=$ 0.96 - 1.08~GeV \cite{Cleymans:2005xv,Cleymans:2006qe}. In \cite{Bravina:2002wz} using a statistical model with two thermal sources (TSM) it was predicted that the average energy per particle as a function of $\sqrt{s_\mathrm{NN}}$ may not be constant, but depend on the resonance share of the produced matter and suggested an increase of $\langle E\rangle/\langle N\rangle$ to 1.1 - 1.2~GeV for $\sqrt{s_\mathrm{NN}}\leq10$~GeV.

\begin{figure} [t!hb]
	\resizebox{0.5\textwidth}{!}{
		\includegraphics{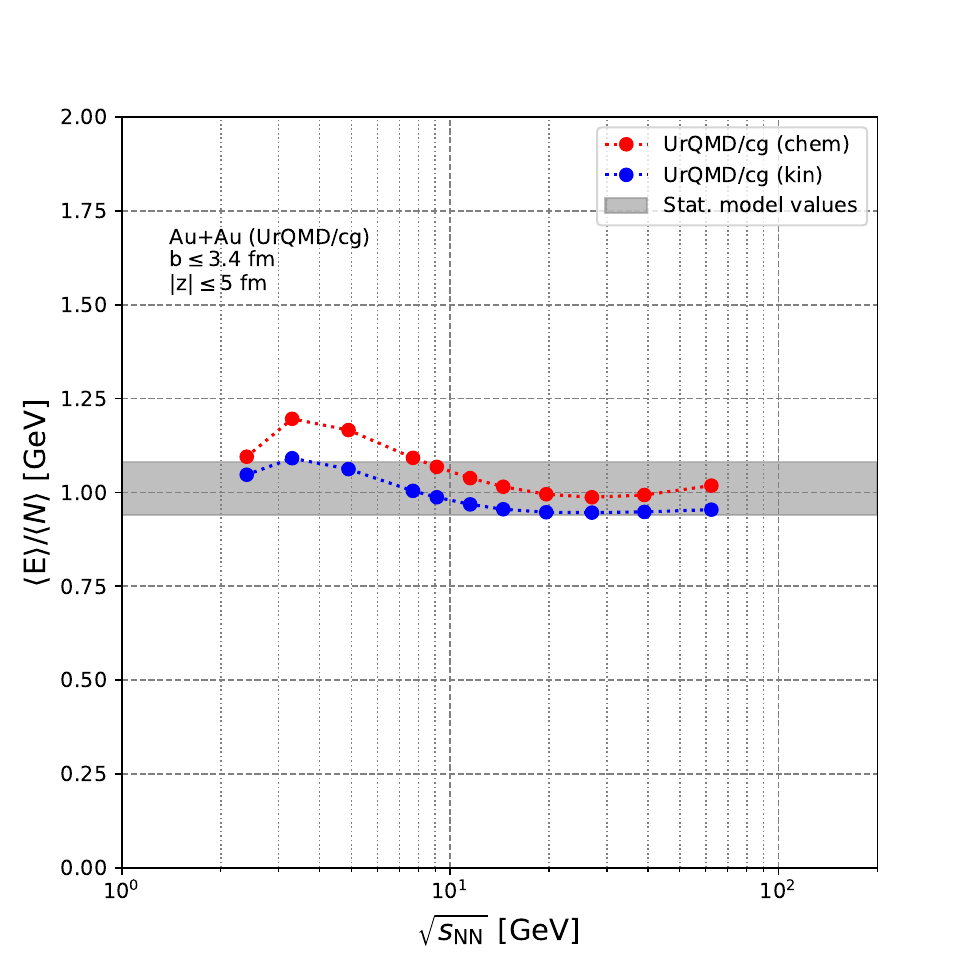}
	}
	\caption{[Color online] The $\langle E\rangle/\langle N\rangle$ at the chemical freeze-out (red circles) and at the kinetic freeze-out (blue circles) extracted from the coarse-graining procedure with a HRG EoS in central in central Au+Au collisions ($b\leq3.4$~fm) from UrQMD. The shaded grey area shows statistical model values ranging from 0.94 to 1.08.}
	\label{EN}
\end{figure}

We show in Fig. \ref{EN} the calculated $\langle E\rangle/\langle N\rangle$ ratio on the chemical and kinetic freeze-out surface arising from the coarse-graining method in dependence of the collision energy. The results at the chemical freeze-out are shown as red circles and the results for the kinetic freeze-out are shown as blue circles. First of all, the computed average energy per particle at the chemical freeze-out hyper surface varies between 0.95 and 1.2 and is therefore very close to the phenomenological 1~GeV/$particle$. In line with \cite{Bravina:2002wz,Cleymans:2006qe}, we observe a slight energy dependence of the ratio with an increase towards 1.2 at the baryon dominated energy region. When studying this ratio on the kinetic freeze-out hyper surface, it is surprising that the same quantity computed here yields only a slightly lower value of $\langle E\rangle/\langle N\rangle\approx0.95$~GeV. This can be explained when examining how $\langle E\rangle/\langle N\rangle$ varies with time which has been done e.g. in \cite{Bleicher:2002dm}. The system is characterized by a plateau in the energy per particle over the duration of the chemical equilibration. Thus we can conclude that the chemical freeze-out line is indeed characterized by $\langle E\rangle/\langle N\rangle\approx1$~GeV/$particle$. Towards lower energies, we predict an increase by about 20~\% in line with \cite{Bravina:2002wz,Cleymans:2006qe}.

\subsubsection{Entropy density}
In thermodynamics, the entropy density is the determining quantity that characterizes when (chemical) equilibrium sets in and is thus a prime candidate to constrain the chemical freeze-out. For an ideal gas of massless hadrons with zero net-baryon density the dimensionless term $s/T^3$ is equivalent to the number of active degrees of freedom. E.g. a pion gas with $s/T^3=3$ would have exactly 3 degrees of freedom. In reality, however, pions are not the only degrees of freedom in the hadron gas, and the entropy has contributions from all mesons and baryons, thus $s/T^3$ is a proxy for the effective degrees of freedom. This relation was extensively studied with the conclusion that chemical equilibration is characterized by a constant entropy density per cubic temperature of 7 (equivalent to 7 effective degrees of freedom) by thermal model studies \cite{CLEYMANS200550}. Therein it was further shown that the decomposition of $s/T^3=7$ into hadronic and mesonic contributions reveals an interchange in the dominant contribution around $\sqrt{s_\mathrm{NN}}=7.7$~GeV. At low energies the degrees of freedom are baryon dominated while at high energies mesonic degrees of freedom are the dominating player.

\begin{figure} [t!hb]
	\resizebox{0.5\textwidth}{!}{
		\includegraphics{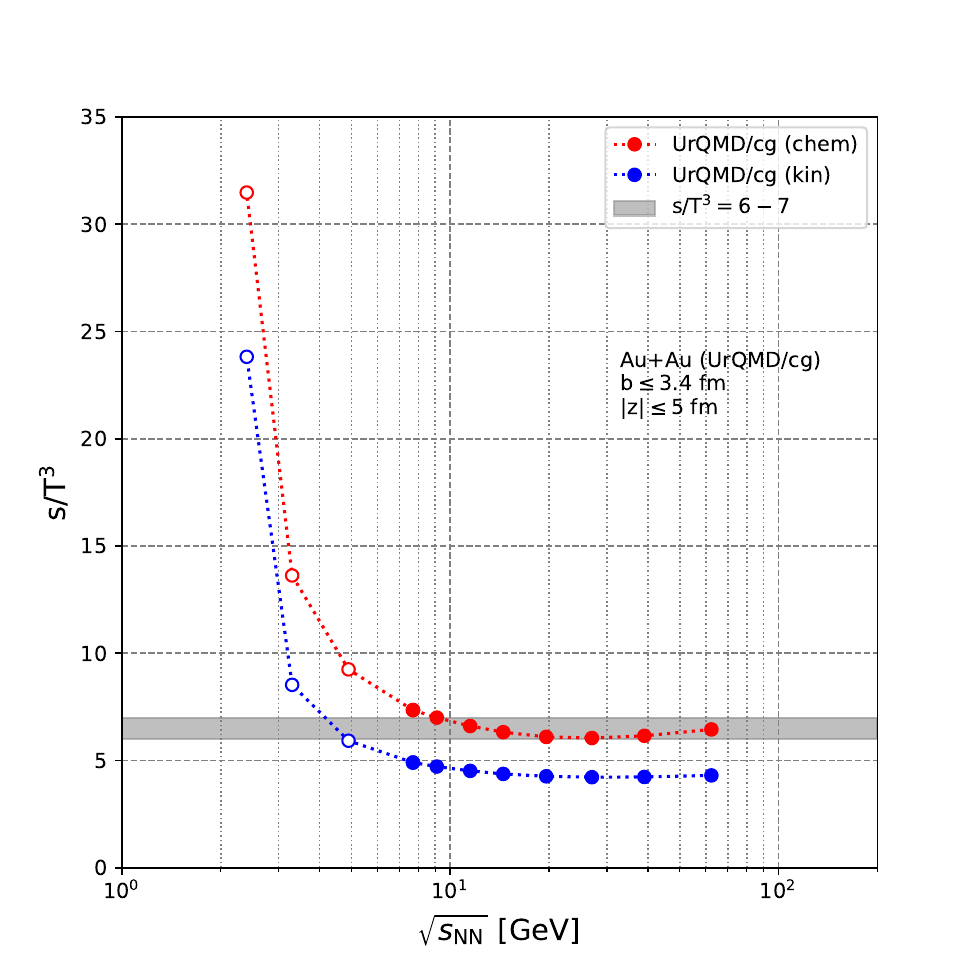}
	}
	\caption{[Color online] The average entropy density $s/T^3$ at the chemical freeze-out (red circles) and at the kinetic freeze-out (blue circles) extracted from the coarse-graining procedure with a HRG EoS in central in central Au+Au collisions ($b\leq3.4$~fm) from UrQMD.}
	\label{sT3}
\end{figure}

The calculated entropy density $s/T^3$ is shown in Fig. \ref{sT3}. The red circles show the values obtained in the present analysis at the chemical freeze-out and the blue circles correspond to the values at the kinetic freeze-out, both in dependence of the collision energy. The filled circles represent the meson dominated region and the empty circles show the baryon dominated region. To avoid numerical problems related to the temperature extraction in cells with high baryon density and low energy density (i.e. cold cells), we include only cells with $T\geq30$~MeV in the averaging.
Extracting the chemical freeze-out from UrQMD, we find that between 7.7~GeV and 62.4~GeV $s/T^3$ remains at values between 6 and 7 lining up with suggested values of Ref. \cite{CLEYMANS200550}. Below the $\sqrt{s_\mathrm{NN}}=8$~GeV, the entropy per cubic temperature rises rapidly as expected from \cite{CLEYMANS200550}. For the kinetic freeze-out, we find that $s/T^3$ saturates between $4-5$ in the meson dominated region and rises in the baryon dominated region.
In conclusion, the $s/T^3$ criterion motivated by the number of degrees of freedom is found to be in line with the previously obtained value of 7 \cite{CLEYMANS200550} in the whole energy range and above $\sqrt{s_\mathrm{NN}}=8$~GeV.

\subsubsection{Baryon and anti-baryon density}
\begin{figure} [t!hb]
	\resizebox{0.5\textwidth}{!}{
		\includegraphics{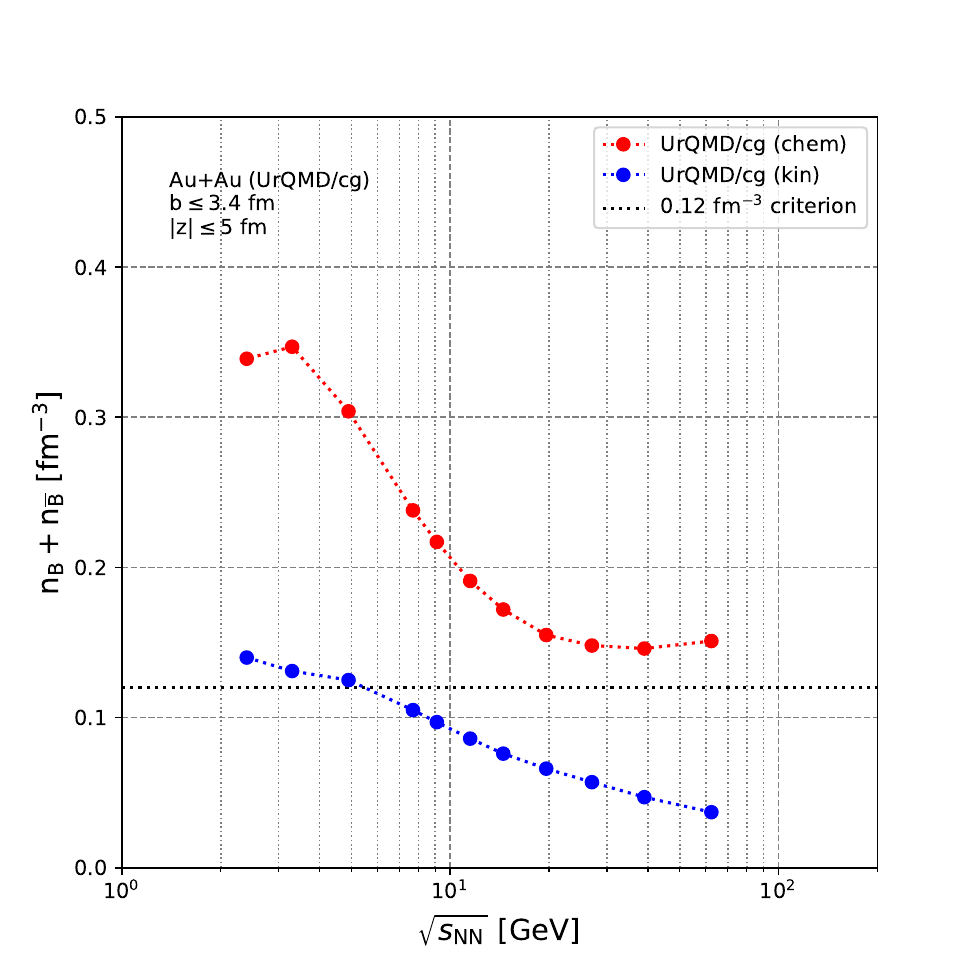}
	}
	\caption{[Color online] The average $n_\mathrm{B}+n_\mathrm{\bar{B}}$ at the chemical freeze-out (red circles) and at the kinetic freeze-out (blue circles) extracted from a coarse-graining procedure with a HRG EoS in central in central Au+Au collisions ($b\leq3.4$~fm) from UrQMD.}
	\label{nB}
\end{figure}

Finally, we explore the suggested baryon-anti-baryon criterion with coarse grained UrQMD. Fig. \ref{nB} shows the calculated $n_\mathrm{B}+n_\mathrm{\bar{B}}$ in dependence of the collision energy. The results for the chemical freeze-out are shown as red circles, while the results for the kinetic freeze-out are shown as blue circles. The results for the total baryon and anti-baryon density at the chemical freeze-out start at a density of $0.35$~fm$^{-3}$ at the low energies and drop rapidly to $0.15$~fm$^{-3}$ at $\sqrt{s_\mathrm{NN}}=20$~GeV. This increase at low energies is qualitatively in line with expectations in \cite{Cleymans:2005xv}. The results at the kinetic freeze-out follow the same trend, but the decrease is more moderate starting at $0.15$~fm$^{-3}$ and ending at $0.05$~fm$^{-3}$. Generally, the baryon density criterion shows a stronger energy dependence than the previously discussed criteria for the chemical freeze-out line. Nevertheless, towards higher energies ($\sqrt{s_\mathrm{NN}}\geq20$~GeV) also this criterion can be applied to characterize the chemical freeze-out line.

\section{Conclusion}
In this article we have developed a novel approach to determine the chemical freeze-out hyper-surface directly from a microscopic simulation. The UrQMD transport model was employed to simulate the underlying events and the microscopic evolution of the system. Using a new algorithm, we traced final state particles back to their original creation space-time coordinate through sequential decay and re-formation chains allowing to reconstruct the chemical freeze-out hyper-surface using a coarse-graining method. We found that the average chemical break-up time remains constant at $\approx7$~fm above $\sqrt{s_\mathrm{NN}}=7.7$~GeV. The extracted temperature and baryon chemical potential values follow the trend of the  established thermal model fits to experimental data. Since UrQMD neither involves a phase-transition to a deconfined state of matter nor the explicit concept of the chemical freeze-out, our results indicate that the chemical freeze-out line may not be indicative of the QCD phase transition, but is defined by the competition of the inelastic scattering rate with the expansion rate. The investigation of previously suggested freeze-out criteria reveals that indeed a constant $s/T^3=7$ and $n_\mathrm{B}+n_\mathrm{\bar{B}}\approx0.12$ fm$^{-3}$ are good proxies to characterize the chemical freeze-out curve at high collision energies ($\sqrt{s_\mathrm{NN}}\geq20$~GeV). The originally suggested Cleymans-Redlich criterion $\langle E\rangle/\langle N\rangle=1$~GeV, however, provides the best estimate for the chemical freeze-out line over all energies.

\section*{Acknowledgments} 
The authors thank Paula Hillmann, Michael Wondrak, Vincent Gaebel, Michel Bonne, Alexander Elz and Harri Niemi for fruitful discussions. G.~Inghirami is supported by the Academy of Finland, Project no. 297058. This work was supported by Deutscher Akademischer Austauschdienst (DAAD), Helmholtz Forschungsakademie Hessen (HFHF) and in the framework of COST Action CA15213 (THOR). The computational resources were provided by the Center for Scientific Computing (CSC) of the Goethe-University Frankfurt. 

\bibliographystyle{epj}
\bibliography{bibliography}
\end{document}